\documentstyle[psfig,aps,pra,twocolumn]{revtex}

\begin{document}
\draft

\title{Bell-inequality violation with ``thermal'' radiation\thanks{This
       paper is dedicated to Professor Jan Pe\v{r}ina in the occasion of
       his $65^{\rm th}$ birthday.}}

\author{Radim Filip,$^1$ Miloslav Du\v{s}ek,$^{2,1}$
Jarom\'\i r Fiur\' a\v sek,$^1$ and Ladislav Mi\v sta$^{1}$}

\address{$^1$Department of Optics, Palack\'y University, 17.~listopadu
   50, 772\,00 Olomouc, Czech~Republic}

\address{$^2$Institute of Physics, Slovak Academy of Sciences,
   D\'{u}bravsk\'{a} cesta 9, 842\,28 Bratislava, Slovakia}

\date{\today}

\maketitle

\begin{abstract}
The model of a quantum-optical device for a conditional preparation
of entangled states from input mixed states is presented. It is
demonstrated that even thermal or pseudo-thermal radiation can be
entangled in such a way, that Bell-inequalities are violated.
\end{abstract}



\section{Introduction}

In last decades, the phenomenon of entanglement between two
spatially separated photons was investigated both experimentally
and theoretically mainly in order to
show that quantum mechanics is not a local realistic theory \cite{micro}.
As a counterpart to the particle-like behavior of photons, the
entanglement of coherent states, that can be considered as
the quantum analogue of deterministic light waves, was examined
\cite{CohSt,Gerry99,nas}. In both the cases, the entangled states, that were
used to test Bell inequalities, were usually considered to be pure
states. Recently, the entanglement of mixed states has been analysed to
understand, how the disorder influences on the amount of entanglement
\cite{mixx}.

In this report, we examine a new situation, when an
entangling device prepares the entangled states of radiation from
mixed states (thermal or pseudo-thermal light) at the input. Similarly
to the idea presented in Ref.~\cite{nas}, the entangling device can
produce a four-mode entangled state with two mixed states and two
vacuum states.
It is shown, that even for very disordered states Bell
inequalities can strongly be violated. If there is a narrow
frequency portion of thermal radiation in the input of the
entangling device then Bell inequalities are violated when the
frequency of radiation is ``low'' and the temperature of thermal
source is ``high''. For a pseudo-thermal radiation the violation of
Bell inequalities is even more significant. In addition,
the violation can be enhanced for both the cases of radiations,
if a lot of  different modes are entangled with vacuum state.
Thus almost the maximal Bell inequality violation can be achieved with
such thermal states exhibiting a large entropy.


\section{Preparation of mixed entangled state}
 \label{device}

We consider two separate systems $A$ and $B$ which consist locally
of two modes $A1$, $A2$ and $B1$ and $B2$. All modes are initially
unentangled. We further assume that the density matrices
of these four modes are diagonal in orthonormal Fock
(number-state) bases $\{ |n\rangle \}$ and that the modes $A2$ and $B2$
are in vacuum states,
\begin{eqnarray}
\hat{\rho}_{A}&=&\sum_{n}p_{n}|n\rangle_{A1}\langle n|\otimes
|0\rangle_{A2}\langle 0|,\nonumber\\
\hat{\rho}_{B}&=&\sum_{m}r_{m}|m\rangle_{B1}\langle m|\otimes
|0\rangle_{B2}\langle 0|.
\end{eqnarray}
The density matrix of the total system has
a factorized form $\hat{\rho}_{\rm in}=\hat{\rho}_{A}
\otimes \hat{\rho}_{B}$. Now, one can consider a conditional
operation which enables to prepare the following
entangled states (for $n\ne 0$ or $m\ne 0$)
\begin{eqnarray}
\label{oper}
|\psi_{nm}\rangle=\frac{1}{\sqrt{2}}\left(
|n\rangle_{A1}|0\rangle_{A2}
|0\rangle_{B1}|m\rangle_{B2}\right.\nonumber\\
\left.-|0\rangle_{A1}|n\rangle_{A2}
|m\rangle_{B1}|0\rangle_{B2}\right).
\end{eqnarray}
The entangling device prepares, for each $m,n$, the analogue of a
singlet state, that was often employed to test Bell-type
inequalities.
Thus the initial density matrix $\hat{\rho}_{\rm in}$ is
transformed into the form
\begin{equation}
\label{final}
\hat{\rho}_{\rm out}= N\sum_{mn}p_{n}r_{m}
(1-\delta_{n0}\delta_{m0})
|\psi_{nm}\rangle\langle \psi_{nm}|,
\end{equation}
where $N=\left[ \sum_{nm} p_n r_m (1-\delta_{n0}\delta_{m0})
\right]^{-1} = (1- p_0 r_0)^{-1}$.

If there is at least one $n>0$ and one $m>0$ such that $p_n\ne 0$ and
$r_m\ne 0$ then state (\ref{final}) is entangled. It can be proved in a
very straightforward way using the so called transposition criterion
\cite{Transp}. This criterion says that if operator
${\hat{\rho}}^{T_B}$, obtained from $\hat{\rho}$ by partial
transposition in subsystem $B$, is not positive the state $\hat{\rho}$
is entangled.
Partial transposition of
$$
  \hat{\rho}_{\rm out} = \sum_{ijklmnst} \!\! \rho_{ijklmnst}\,
              |i_{A1} j_{A2} k_{B1} l_{B2} \rangle  \langle
               m_{A1} n_{A2} s_{B1} t_{B2}|
$$
in basis $|i_{A1} j_{A2} k_{B1} l_{B2} \rangle \equiv
|i\rangle_{A1}|j\rangle_{A2} |k\rangle_{B1}|l\rangle_{B2}$ gives
$$
  \hat{\rho}_{\rm out}^{T_B} = \sum_{ijklmnst} \!\! \rho_{ijklmnst}\,
              |i_{A1} j_{A2} s_{B1} t_{B2} \rangle  \langle
               m_{A1} n_{A2} k_{B1} l_{B2}|.
$$
Now, let us suppose vector
\begin{eqnarray*}
  |\phi_{mn}\rangle=\frac{1}{\sqrt{2}}\left(
  |0\rangle_{A1}|m\rangle_{A2}
  |0\rangle_{B1}|n\rangle_{B2}\right. \\
  \left.+|m\rangle_{A1}|0\rangle_{A2}
         |n\rangle_{B1}|0\rangle_{B2}\right)
\end{eqnarray*}
where $m,n>0$ and calculate the following mean value
\begin{equation}
 \langle \phi_{mn} | \,
  {\hat{\rho}}_{\rm out}^{T_B}
  \, | \phi_{mn}\rangle
  = - N {p_m r_n \over 2}.
\label{RedCr}
\end{equation}
If $p_m \ne 0$ and $r_n \ne 0$,
this quantity is {\it negative}.
On the other hand, the entanglement of discussed states can often be
``masked'' by the noise of original mixed states. E.g., conditional
von Neumann entropy, $S({\hat{\rho}}'_A)-S(\hat{\rho})$, is positive
for many particular cases here. Nevertheless, we will show that the
entanglement is ``strong'' enough to violate CHSH-Bell inequality.

\begin{figure}
\medskip
\centerline{\psfig{width=0.9\hsize,file=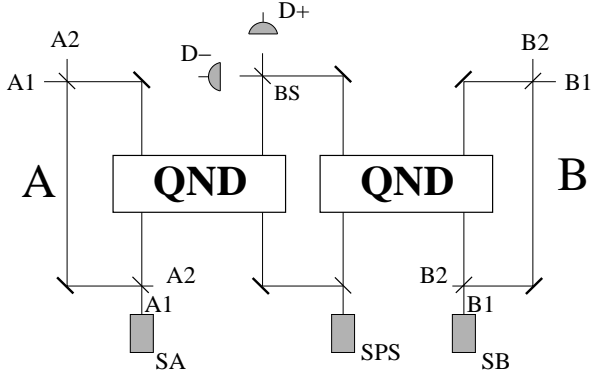,clip=}}
\medskip
\caption{Preparation device; $SA$ and $SB$ denote the sources of thermal
(pseudothermal) radiation, $SPS$ is a single photon source, $QND$ is
quantum nondemolition measurement performed by the Kerr interaction
and $D_{+}$ and $D_{-}$ are detectors.}
\label{fig1}
\end{figure}

The proposed conditional operation can be, in principle, realized
in the following way (see Fig.~\ref{fig1}) \cite{nas}: Let us assume a
Mach-Zehnder (M-Z) interferometer with equal-length arms and with
one photon in its input.
Into both the arms of the interferometer we insert a nonlinear Kerr
medium effectively described by the following interaction Hamiltonian
\begin{equation}
\hat{H}_{I,i}=\hbar\kappa\hat{a}^{\dag}\hat{a}\hat{a}_{i1}^{\dag}\hat{a}_{i1},
\end{equation}
where $\hat{a}^{\dag}$ and $\hat{a}$ are the creation and annihilation
operators of the mode corresponding to the left (or right) arm in the
M-Z interferometer, $\hat{a}_{i1}^{\dag}$ and
$\hat{a}_{i1}$, with $i=A,B$, are the creation
and annihilation operators of modes
$A1$ (or $B1$), and $\kappa$ is a real interaction constant.

If there is a photon in the left arm of the central M-Z interferometer and
the product $\kappa\tau_{\rm int}$, where $\tau_{\rm
int}$ is an effective interaction time,\footnote{Quantity $\tau_{\rm
int}$ has the meaning of the parameter of the device. It is not a
usual time variable. It represents the effective expression
of the fact that the nonlinear medium has finite dimensions.}
is set to be equal exactly to $\pi$ then the
described device realizes the phase shift $\pi$ in the left M-Z
interferometer, $A$, and effectively flips the modes $A1$ and $A2$ on the
output. On the other hand, if there is no photon in the left arm then
the states of modes $A1$ and $A2$ stay unchanged,
\begin{eqnarray}
\hat{U}_{A}|n\rangle_{A1}|0\rangle_{A2}|1\rangle&=&|0\rangle_{A1}|n\rangle_{A2}|1\rangle,\nonumber\\
\hat{U}_{A}|n\rangle_{A1}|0\rangle_{A2}|0\rangle&=&|n\rangle_{A1}|0\rangle_{A2}|0\rangle.
\end{eqnarray}
The same is true about the right arm of the central M-Z interferometer and
modes $B1$ and $B2$.
These unitary transformations $\hat{U}_{i}$, $i=A,B$, can be expressed as
\begin{equation}
\hat{U}_{i}=\hat{U}_{BS,i}^\dagger\hat{U}_{I,i}\hat{U}_{BS,i},
\end{equation}
where $\hat{U}_{BS,i}$ is the $50:50$ beam splitter transformation
and $\hat{U}_{I,i}$ accounts for the nonlinear interaction in Kerr medium,
\begin{eqnarray}
\hat{U}_{BS,i}&=&\exp\left[\frac{\pi}{4}(\hat{a}_{i2}^{\dag}\hat{a}_{i1}
-\hat{a}_{i1}^{\dag}\hat{a}_{i2})\right], \nonumber\\
\hat{U}_{I,i}&=&\exp(i\pi\hat{a}^{\dag}\hat{a}\hat{a}_{i1}^{\dag}\hat{a}_{i1}).
\end{eqnarray}
So, if the photon goes through
the left arm the modes $A1$ and $A2$ are flipped while
the state of system $B$ is unchanged. Completely symmetrical
situation occurs, if the photon goes through the right arm.

Due to the path uncertainty of the photon in the interferometer
the state of the whole system after the interaction is given by the
formula
\begin{eqnarray}
|\Psi\rangle=\frac{1}{\sqrt{2}}(|0\rangle|1\rangle\,
|n\rangle_{A1}|0\rangle_{A2}
|0\rangle_{B1}|m\rangle_{B2}+\nonumber\\
+i|1\rangle|0\rangle\,
|0\rangle_{A1}|n\rangle_{A2}
|m\rangle_{B1}|0\rangle_{B2}),
\end{eqnarray}
where the kets without any subscript denote possible states of the
photon inside the M-Z interferometer situated in the center.
Which-way information is finally erased \cite{eraser} by a beam splitter
with amplitude reflectivity $i/\sqrt{2}$ (the last one in the M-Z
interferometer) followed by two photodetectors $D_{+}$ and $D_{-}$
(see Fig.~\ref{fig1}).
Depending on which one of these two detectors fires
we obtain one of two possible output states of modes $A1$, $A2$,
$B1$, and $B2$. Detector $D_{+}$ fires with probability
$w_{+}=(1+\delta_{n0}\delta_{m0})/2$ and if it clicks the following
state is obtained
\begin{eqnarray}
|\Psi_{+}\rangle&=&\frac{1}{\sqrt{2}}(
|n\rangle_{A1}|0\rangle_{A2} |0\rangle_{B1}|m\rangle_{B2}
\nonumber\\
&&+ |0\rangle_{A1}|n\rangle_{A2}
|m\rangle_{B1}|0\rangle_{B2}.
\end{eqnarray}
Similarly, detector $D_{-}$ clicks with probability
$w_{-}=(1-\delta_{n0}\delta_{m0})/2$ and when it fires one
obtains the state
\begin{eqnarray}
|\Psi_{-}\rangle&=&\frac{1}{\sqrt{2}}(
|0\rangle_{A1}|n\rangle_{A2}
|m\rangle_{B1}|0\rangle_{B2},
\nonumber\\
&&-|n\rangle_{A1}|0\rangle_{A2}
|0\rangle_{B1}|m\rangle_{B2}
\end{eqnarray}
which is exactly the considered state (\ref{oper}).


\section{Bell-inequality violation}

In order to demonstrate the violation of Bell inequalities one needs
local operations analogous to spin rotations. In our particular case
the following operations do the job
\begin{eqnarray}
|n\rangle_{1}|0\rangle_{2} &\rightarrow&
  \cos\theta\,
  |n\rangle_{1}|0\rangle_{2} + \sin\theta\,
  |0\rangle_{1}|n\rangle_{2}
  \mbox{~for~} n \ne 0,\nonumber\\
|0\rangle_{1}|n\rangle_{2}&\rightarrow&
  -\sin\theta\,
  |n\rangle_{1}|0\rangle_{2} + \cos\theta\,
  |0\rangle_{1}|n\rangle_{2}
  \mbox{~for~} n \ne 0, \nonumber\\
|0\rangle_{1}|0\rangle_{2}&\rightarrow& |0\rangle_{1}|0\rangle_{2},
  \label{operace}
\end{eqnarray}
where $\theta$ is the parameter of transformation, it does not
depend on $n$.

Bell-type experiment consists of two ``rotations'' according to
recipe (\ref{operace}), performed by two possibly space-like
separated observers, followed by realistic yes--no detection on each
mode. Each such detection has only two possible outcomes (detector
either fires or it does not), that can be described by projectors
$|0\rangle\langle 0|$ (for ``no'') and
$\hat{1}-|0\rangle\langle 0|=\sum_{n=1}^{\infty}|n\rangle\langle n|$
(for ``yes'').
Let us assign the following values to these outcomes: $z_i$=0
if the detector (in mode $i$) is quiet and $z_i$=1 if it clicks.
Then the results $X$ and $Y$ of local two-mode measurements
(including ``rotations'') performed by the first and the second
observer, respectively, can be expressed as
\begin{eqnarray}
 \label{observation}
 X(\theta) &=&
 z_{A1}(\theta)-z_{A2}(\theta), \nonumber \\
 Y(\theta) &=&
 z_{B1}(\theta)-z_{B2}(\theta).
 \end{eqnarray}
After the experiment is repeated many times and our two observers
compare their results, the mean value of Bell operator (for CHSH
inequalities) can be estimated,
\begin{equation}
\label{def-B}
{\cal B}
=|C(\theta_A,\theta_B)+C(\theta_A,\theta'_B)+C(\theta'_A,\theta _ B)
-C(\theta'_A,\theta'_B)|,
\end{equation}
where correlation function
\begin{equation}
C(\theta_1,\theta_2) \equiv
\sum_{j,k} X_j Y_k\, p(X_j,Y_k|\theta_A,\theta_B)
\label{corr}
\end{equation}
(summations go over all possible results).
Every local-realistic theory \cite{EPR} must fulfill the following
inequality ${\cal B} \le 2$ \cite{Bell}. However, it follows from
straightforward quantum-mechanical calculations that
for state (\ref{final}) the correlation function (\ref{corr}) reads
\begin{equation}
\label{corr2}
C(\theta_{A},\theta_{B})=-\cos\left[2(\theta_{A} - \theta_{B})\right]
\frac{(1-p_0)(1-r_0)}{1-p_0 r_0}.
\end{equation}
Therefore the results of the above mentioned local measurements
performed on state (\ref{final}) {\it violate\/} inequality ${\cal B}
\le 2$ in principial. Maximal violation,
\begin{equation}\label{max}
{\cal B}_{\rm max}=2\sqrt{2} \, \frac{(1-p_0)(1-r_0)}{1-p_0 r_0},
\end{equation}
occurs for the angles
\begin{equation}
\label{uhly}
\theta_{A}=0,~~\theta'_{A}=\frac{\pi}{4}~~
\theta_{B}=\frac{\pi}{8},~~\theta'_{B}=-\frac{\pi}{8}.
\end{equation}
If both the mixed states have the same overlap with vacuum state
$p_{0}=r_{0}$, the condition for the violation of
Bell inequality for the considered angles is given in a simple form
\begin{equation}
p_{0}<\frac{\sqrt{2}-1}{\sqrt{2}+1}\approx 0.1716.
\end{equation}
As one can see the maximum value of ${\cal B}$ depends on the
probability of the presence of the vacuum state in the input density
matrices. Thus, if the input density matrices of systems $A1$ and
$B1$ do not contain the vacuum state the maximal violation of
CHSH-Bell inequality is the same as for the pure EPR maximally
entangled state of two spin-half particles. In the opposite case, the
mean value of Bell operator decreases as the contribution of the
vacuum state increases in the mixtures. It should be noticed
that for properly chosen local measurements the violation of
CHSH-Bell inequality does not depend on the randomness contained in
the mixture but only on the overlaps of the vacuum state and the
input density matrices.


\section{Thermal and pseudo-thermal radiation}

There are two mixed states of special interest, namely
thermal radiation, exhibiting Bose-Einstein statistics, and
pseudo-thermal radiation, exhibiting Poissonian statistics.
Let us study now the entangled states prepared by the device
proposed in Sec.~\ref{device} when thermal and pseudo-thermal states
are at the input.

A single mode of thermal radiation has the density matrix
\begin{equation}
 \hat{\rho}=\sum_{n}\frac{\langle
 n\rangle^{n}}{(1+\langle n\rangle)^{1+n}} \,|n\rangle\langle n|,
\end{equation}
where
\begin{equation}
\langle n\rangle=
\frac{1}{\exp(\frac{\hbar\omega}{k_{B}T})-1}.
\end{equation}
For example, if
the temperature of a radiation source (e.g., incandescent
lamp) $T\approx 3000\,$K  and the optical frequency
$\omega\approx 2.5\times 10^{15}\,$Hz, the mean value of photon
number is $\langle n\rangle\approx 1.77\times 10^{-3}$.
The probability of the vacuum state in the mixture is
\begin{equation}
p_{0}=1-\exp\left(-\frac{\hbar\omega}{k_{B}T}\right)=\frac{1}{1+\langle
n\rangle},
\end{equation}
what leads to the value
$p_0\approx 0.9982$ for the above given data.
Thus in the optical region, the
overlap of vacuum and thermal light is too large and the
Bell-inequality violation does not occur.

\begin{figure}
\medskip
\centerline{\psfig{width=0.9\hsize,file=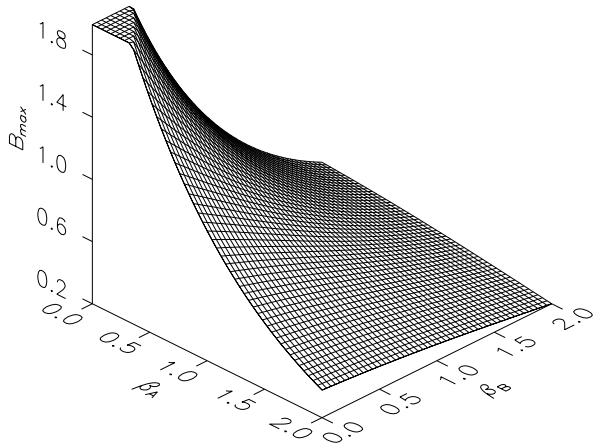,clip=}}
\medskip
\caption{The maximal violation of the CHSH-Bell inequality for
thermal light as the function of parameters
$\beta_{A}=\hbar\omega_{A}/k_{B}T_{A}$ and
$\beta_{B}=\hbar\omega_{B}/k_{B}T_{B}$}
\label{fig2}
\end{figure}

The dependence of the
maximal Bell-inequality violation on the parameter
$\beta_{i}=\hbar\omega_{i}/k_{B}T_{i}$, $i=A,B$
of particular modes $A1$, $B1$
can be simply evaluated:
\begin{eqnarray}
{\cal B}_{\rm
max}&=&2\sqrt{2}\frac{1}{\exp(\beta_{A})+\exp(\beta_{B})-1}=\nonumber\\
& &=2\sqrt{2}\frac{1}{1+\langle n\rangle_{A}^{-1}+\langle n\rangle_{B}^{-1}}
\end{eqnarray}
and
it is displayed in Fig.~\ref{fig2}. Only for very small $\beta_{A}$
and $\beta_{B}$, i.e., for high temperatures and small frequencies,
CHSH-Bell inequality is violated. Thus for the given temperature $T$
of both the thermal sources the infrared component of radiation
gives better results than the ultra-violate one. On the other hand,
for the fixed frequency $\omega$ of both the sources the higher
temperature leads to the stronger violation of Bell inequality. If
both the sources are identical the Bell-inequality violation occurs
only if the dimensionless parameter $\beta$ satisfies relation
$\beta<\ln\frac{\sqrt{2}+1}{2}\approx 0.1882$
or the mean number $\langle n\rangle$
is sufficiently large $\langle n\rangle> 2(\sqrt{2}+1)\approx 4.828$.
Consequently, for the visible component of
radiation the thermal sources must have an ``astronomical'' temperature
$T>101000\mbox{K}$, whereas for the infrared component with
$\omega\approx 5\times 10^{13}\mbox{Hz}$, temperature
$T>2021\mbox{K}$ is sufficient to obtain Bell-inequality violation.

\begin{figure}
\medskip
\centerline{\psfig{width=0.9\hsize,file=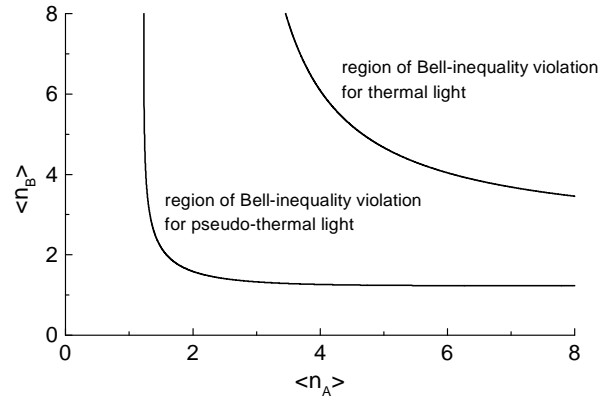,clip=}}
\medskip
\caption{The border of violation of CHSH-Bell inequality for thermal
and pseudo-thermal light
in dependence on mean photon numbers $\langle n\rangle_{A}$
and $\langle n\rangle_{B}$.}
\label{fig3}
\end{figure}

Another interesting kind of mixed state is that corresponding to
pseudo-thermal light \cite{pseudoth}. Its density matrix can be written as
\begin{equation}\label{pseudo}
\hat{\rho}=\sum_{n}\frac{\langle n\rangle^{n}}{n!}e^{-\langle
n\rangle}|n\rangle\langle n|.
\end{equation}
Pseudo-thermal radiation can be obtained from an intensity-stabilized
single mode laser with the phase uniformly distributed in the
interval $\langle 0,2\pi)$.
In contrast to thermal radiation, the
maximally probable state in the mixture (\ref{pseudo}) is not vacuum
state but it is state $|n\rangle$, where $n$
corresponds approximately to the mean number of photons $\langle n
\rangle$. Thus the overlap of pseudo-thermal light with the vacuum
state is much less than for thermal light. The probability of the
vacuum state in the density matrix (\ref{pseudo}) is $p_{0}=\exp(-\langle
n\rangle)$. This leads to maximal Bell-inequality violation
\begin{equation}
{\cal B}_{\rm max}=2\sqrt{2}\frac{[1-\exp(-\langle
n\rangle_{A})][1-\exp(-\langle n\rangle_{B})]}{1-\exp[-(\langle
n\rangle_{A}+\langle n\rangle_{B})]}.
\end{equation}
From Fig.~\ref{fig3} one can see that in the case of
pseudo-thermal light the Bell-inequality violation is achieved for
less $\langle n\rangle_{A}$ and $\langle n\rangle_{B}$ than in
the case of thermal light. If one considers two identical pseudothermal
sources, then the Bell inequality is violated if $
\langle n\rangle>\ln\frac{\sqrt{2}+1}{\sqrt{2}-1}$.
As the pseudo-thermal light with such a mean photon number can
be experimentally achieved from the laser light in optical frequencies,
the violation can be obtained more simply than for the thermal light.

Real light sources emit to a large amount of different independent modes.
The density matrix of this multi-mode state is given
in the following form
\begin{equation}\label{multi}
 \hat{\rho}=\prod_{\mu}\,\,\sum_{n_{\mu}=0}^{\infty}\frac{\langle
 n_{\mu}\rangle^{n_{\mu}}}{(1+\langle n_{\mu}\rangle)^{1+n_{\mu}}}
 \,|n_{\mu}\rangle\langle n_{\mu}|,
\end{equation}
where $n_{\mu}$ is photon number for particular mode $\mu$ and
$|n_{\mu}\rangle$ is the Fock state of the corresponding mode.
Let us suppose that this multi-mode
thermal state is feeded to the inputs $A1$ and $B1$ and the multi-mode
vacuum states are present in the inputs $A2$ and
$B2$.\footnote{In reality there could be a problem to set the proper
    parameters of Kerr interaction for all the frequancy components
    together.}
The analysis presented in Sec. III may be
generalized to multi-mode light in a straightforward way. We define the
``rotations'' of the multi-mode vacuum $|\{0\}\rangle$ and any excited
multi-mode state $|\{n\}\rangle$ as follows,
\begin{eqnarray}
|\{n\}\rangle_{1}|\{0\}\rangle_{2} &\rightarrow&
  \cos\theta\,
  |\{n\}\rangle_{1}|\{0\}\rangle_{2} + \sin\theta\,
  |\{0\}\rangle_{1}|\{n\}\rangle_{2},\nonumber\\
|\{0\}\rangle_{1}|\{n\}\rangle_{2}&\rightarrow&
  -\sin\theta\,
  |\{n\}\rangle_{1}|\{0\}\rangle_{2} + \cos\theta\,
  |\{0\}\rangle_{1}|\{n\}\rangle_{2},\nonumber\\
\end{eqnarray}
for $\{n\}\not=0$, and for multi-mode vacuum in both the modes:
$|\{0\}\rangle_{1}|\{0\}\rangle_{2}\rightarrow
|\{0\}\rangle_{1}|\{0\}\rangle_{2}$.
Detection that discriminates between the field
vacuum and other states has two possible outcomes described by
projectors  $|\{0\}\rangle\langle\{0\}|$ and
$\hat{1}-|\{0\}\rangle\langle\{0\}|$.
It can be shown that the maximal violation of Bell inequality exhibits
the same form (\ref{max}) as in the case of single-mode radiation,
but with  the following notation
\begin{equation}
p_{0}=\prod_{\mu}p_{0,\mu},\hspace{0.2cm}r_{0}=\prod_{\mu}r_{0,\mu}.
\end{equation}
With increasing number of the modes of thermal radiation
the effective overlap of vacuum state and such a multi-mode field
decreases and, consequently, the maximal violation of Bell
inequality is enhanced. In this way, the Bell inequality violation can
be achieved for every thermal radiation, if the sufficient number of
modes is taken into account.


\section{Conclusion}

An entangling device employing non-linear dynamics and postselection
has been proposed and it has been shown that two mixed states can be
entangled in such a way that the entanglement of the resulting state
is strong enough to violate Bell inequalities (when proper local
measurements are chosen). The disorder due to the statistical nature
of the density matrices of input states is irrelevant -- it does not
influence the violation of Bell inequality. The only parameters
affecting the maximum of the mean value of Bell operator are overlaps
$p_{0}=\langle 0|\rho_{A1}|0\rangle$ and $r_{0}=\langle 0|\rho_{B1}|0\rangle$.
This is also the reason of a contra-intuitive behavior when the
entanglement increases as the input thermal state becomes more
`classical' ($\beta\rightarrow 0$), whereas in the `quantum' limit
($\beta\rightarrow\infty$) the entanglement vanishes.
Another contra-intuitive aspect of this phenomena appears if the
multi-mode thermal radiation is considered. Since the overlap with
multimode vacuum becomes smaller as the number of modes increases,
the multi-mode thermal radiation can violate
Bell inequality more notably,
irrespective of its larger entropy. Thus this ``classical-like''
radiation can be strongly entangled in
the ideal case and even exhibit the pronounced quantum nonlocality.
Unfortunately, like the other kinds of mesoscopic states, the described
quantum superpositions are very sensitive to the descructive influence of
decoherence and losses.

\section*{Acknowledgments}

This research was supported under the project LN00A015 of the
Ministry of Education of the Czech Republic and the European Union project
EQUIP (contract IST-1999-11053). R.F. thanks to
J. \v Reh\' a\v cek for stimulating discussions.


\end{document}